\documentclass[ final  ] {aipproc}

\layoutstyle{8x11single}

\begin{document}

\title{Energy-time entangled qutrits: \\  Bell tests and quantum communication}

\author{R.~T.~Thew}{  address={Group of Applied Physics, University of Geneva, 1211 Geneva 4, Switzerland}}

\author{S.~Tanzilli}{
  address={Group of Applied Physics, University of Geneva, 1211 Geneva 4, Switzerland}}

\author{A.~Ac\'in}{ address={Institut de Ci\`encies Fot\`oniques, Jordi Girona 29, 08034 Barcelona, Spain}}

\author{H.~Zbinden}{
  address={Group of Applied Physics, University of Geneva, 1211 Geneva 4, Switzerland}}

\author{N.~Gisin}{
  address={Group of Applied Physics, University of Geneva, 1211 Geneva 4, Switzerland}}

\begin{abstract}
  We have developed a scheme to generate, control, transmit and measure entangled photonic qutrits (two photons each of dimension $d = 3$).  A Bell test of this source has previously been reported elsewhere \cite{Thew04b}, therefore, here we focus on how the control of the system is realized. Motivated by these results, we outline how the scheme can be used for two specific quantum protocols, namely  key distribution and coin tossing and discuss some of their advantages and disadvantages.  
\end{abstract}

\maketitle



Performing quantum communication with high-dimensional systems would appear to be an obvious and straight-forward extension to many of the qubit protocols that have driven quantum information science in recent years. There have been theoretical proposals for Quantum Key Distribution (QKD)  \cite{Bechmann00a} with greater  security \cite{Cerf02a} as well as specific proposals for qutrits such as quantum coin tossing  \cite{Ambainis02a}. Fundamentally, high-dimensional Bell inequalities  have revealed greater violations of non-locality \cite{Collins02a} while increasing the dimensions of the entangled systems has been shown to facilitate closing the detection loop-hole \cite{Massar02a} for such tests. Optics has been able to provide a realistic test-bed for these proposals with several different schemes being realized for generating high-dimensional entanglement \cite{QutritsExp, Langford04a} as well as the scheme presented  here \cite{Thew04b,Thew04a}.


This source of  entangled qutrits is  analogous to the energy-time qubit arrangement of Franson \cite{Franson89a}. The experimental scheme has been developed for telecom wavelengths and with proven long distance quantum communication architecture \cite{Tittel99a,Thew02a} to optimise the usefulness of this high dimensional entanglement resource. We have already performed a Bell-type test on this system \cite{Thew04b} where we determined a means of inferring a violation from the interference fringe visibility. We found  a net visibility of V$_{net}$= $0.979 \pm 0.006$ corresponding to a violation of the V$_{Bell}$ limit by 34$\,\sigma$. From a practical perspective, it is interesting to discuss how the symmetry constraints were met and controlled in this set-up to achieve the violation. In doing this we also explore some of the subspaces of these high-dimensional states and we show how these correspond to the states required for quantum coin tossing. We also outline how we can use this scheme for QKD. We also briefly highlight some of the positive as well as negative aspects of these higher dimensional states in the case of these  proposed protocols.



\begin{figure}[h!]\label{fig:expschematic}
\includegraphics[width=11cm,height=3.9cm]{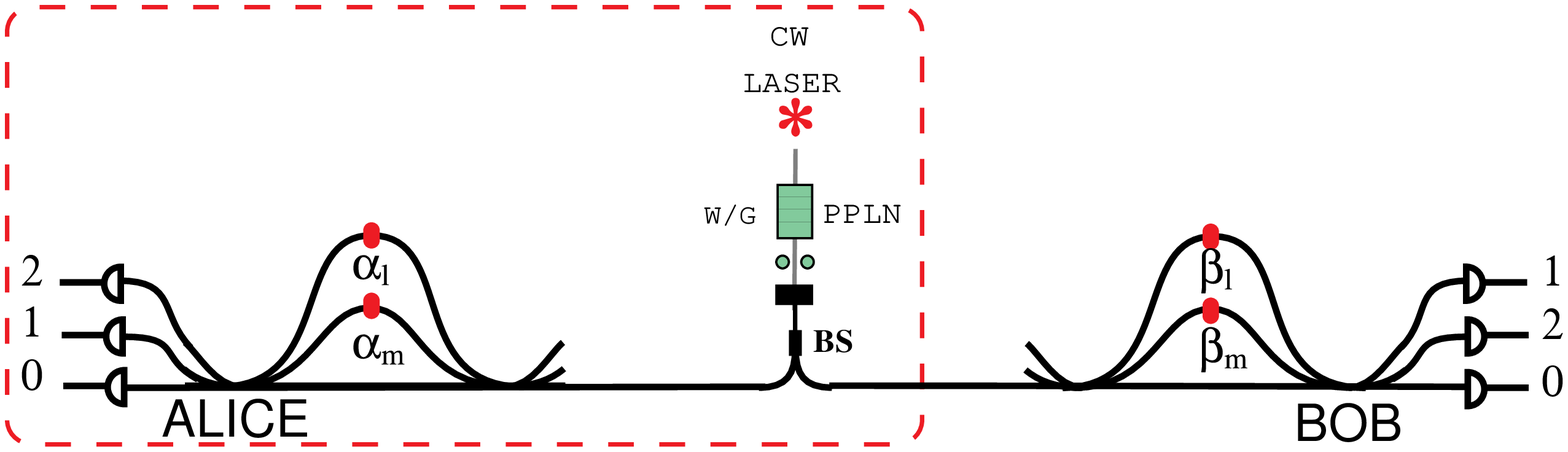}
\includegraphics[width=5cm,height=3.9cm]{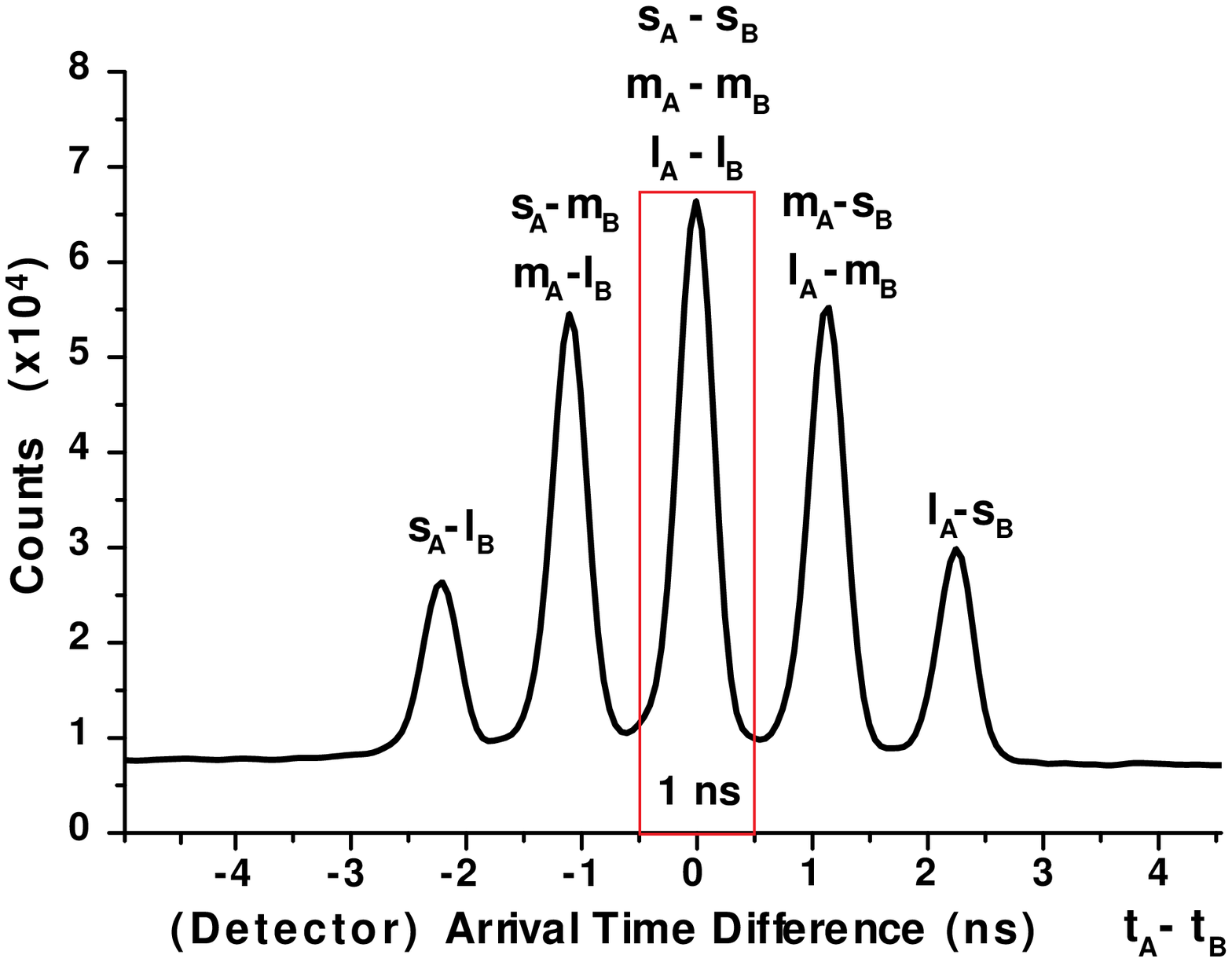}
\caption{(left) Schematic of experimental arrangement for energy-time entangled qutrits. (right) Histogram of the photon arrival time difference at pairs of  Alice and Bob's detectors. The labeling denotes the possible interferometer-path combinations.}
\end{figure}

The experimental set-up is illustrated in the schematic of Fig.\ref{fig:expschematic}. We use energy-time entangled photon pairs created at telecom wavelengths, via a PPLN waveguide (courtesy of the Uni. of Nice) \cite{Tanzilli01a}, and two three-arm interferometers  to generate and analyze entangled qutrits. For each interferometer we can define a phase vector consisting of the two independent phases, e.g. the relative phases between the short-medium (m) and short-long (l), path-lengths. Coincidence measurements at the outputs of the interferometers project onto entangled qutrit states defined when the photons take the same path, i.e.  short-short,  medium-medium or long-long  in each of  Alice and Bob's interferometers. 
In complete analogy with the Franson qubit arrangement \cite{Franson89a}, we use a continuous wave (CW) laser  where the long coherence length does not provide a well defined creation time for the photon pairs. The coherence length of the photon pairs is much longer than the path-length differences in the interferometers, while the coherence length of the single photons is much less,  such that no single-photon interference takes place. We have set the path-length differences, $l-m \approx m-s$, corresponding to a time difference of 1.2\,ns. An arrival-time-difference histogram with five peaks, due to all the  possible  path combinations, like that  on the right in Fig.\ref{fig:expschematic}  is generated for each detector combination. Therefore, when we select events in the central peak, i.e.. $ \Delta t = t_A -t_B \approx 0$,  these correspond to the  post-selection of states of the form
\begin{eqnarray}
|\psi_{j,k}\rangle &\propto & c_s|ss\rangle + c_me^{i (\alpha_m + \beta_m + \chi_{jk}^m)}|mm\rangle  +  c_le^{i(\alpha_l + \beta_l + \chi_{jk}^l)}|ll
\rangle . \label{eq:qutrit}
\end{eqnarray}
Here $\alpha_m, \alpha_l$ and $\beta_m,  \beta_l$, represent the phases in Alice and Bob's medium and long interferometer arms. 
 $\chi_{jk}^m$ and $\chi_{jk}^l$ are multiples of $2\pi/3 $ due to the symmetry of the 3x3 couplers and  depends on the paths taken by the photons in the interferometers and which output (detector), $j,k \in \{0,1,2\}_{A,B}$. For more technical as well as theoretical details see refs \cite{Thew04b,Thew04a,Zukowski97a}. Maximally entangled qutrits require that  $|c_s|^2 = |c_m|^2 = |c_l|^2$, which simply corresponds to  having symmetric splitting ratios for the fiber couplers in the interferometers. We can then observe the  three orthogonal states corresponding to the three different coincidence detections, $0_A0_B$, $  0_A1_B$, $0_A2_B$, or  their cyclic permutations. When the phases are scanned the coincidences vary as a function of Alice and Bob's phase vectors.



The first question that arises now is - how do we characterise the set-up? Being an interferometric arrangement the most obvious solution is to vary the phases and observe the interference fringe visibility, the normalized difference between the maximum and minimum intensities/coincidence rates. Classically, the visibility is defined by  $ V =(I_{max}-I_{min})/(I_{max}+I_{min})$. To obtain a maximum of 1 we need $I_{min}= 0$. This is theoretically trivial in the case  of qubits with only one phase.  With two phases, as in the qutrit scenario, we can always set one of the phases such that changing the other never results in a minimum of zero. Therefore, there are constraints on how the two phases must be related to obtain the maximum visibility. Practically this presents a problem since the phase is controlled here via temperature and thus any variation in temperature implies a change in the phase. This means that a phase of  "0" at the beginning of the experimental run may not be "0" at the end due to temperature drift. We also generally scan through all phase settings, thus, we need a means of determining how the two phases are varying in relation to one another. 

To this end we revisit the histogram of five peaks in Fig.\ref{fig:expschematic}. The two peaks either side of the central peak at $ \Delta t = t_A -t_B = \pm 1.2\,ns$, the " {\it left}" and the " {\it right}", provide us with a control solution.  This time difference is defined by the  path-length differences between the short-medium and medium-long arms of the interferometers. These {\it satellite} peaks correspond to projections onto sub-spaces of the entangled qutrit Hilbert space. For example, the $0_A0_B$ detector combination has the unnormalised states for the  {\it left}, $|\psi_{00}^L\rangle $, and {\it right},  $|\psi_{00}^R\rangle$, peaks  (for simplicity we set $\beta_j =0$): 
\begin{eqnarray}
 |\psi_{00}^L\rangle \propto  |ms\rangle + e^{i(\Phi_{00}^L- 2\pi/3)}|lm\rangle
    \hspace{2cm}   |\psi_{00}^R\rangle \propto |sm\rangle + e^{i(\Phi_{00}^R+2\pi/3)}|ml\rangle
\label{eqLRphasestates}.
\end{eqnarray}
By defining the phases as $\Phi_{00}^L \equiv \alpha_{l}-\alpha_{m}$ and $\Phi_{00}^R \equiv \alpha_{m}+ 2\pi/3$,  we can rewrite the states of the central peak as, $ |\psi_{00}\rangle \propto  |00\rangle + \exp[i\Phi_{00}^R]|11\rangle +  \exp[i(\Phi_{00}^R+\Phi_{00}^L)]|22 \rangle $,
and in general, for all detector combinations, we find the probabilities for coincidence detection given by
\begin{eqnarray}
   P_{jk} \propto 3 + 2\lambda \left[\cos(\Phi_{jk}^R) + \cos(\Phi_{jk}^L) + \cos(\Phi_{jk}^R+\Phi_{jk}^L)\right].\label{eq:fringes}
\end{eqnarray}
This is actually the probability associated with the mixed state, $\rho = \lambda |\psi_{jk}\rangle \langle \psi_{jk}| + (1- \lambda)I/9$, where $I$ denotes symmetric noise \cite{Thew04b}. The requirements for the minimum are the same, however the mixing parameter will also effect this minimum. Therefore, by monitoring the way the states associated with the   {\it left} and  {\it right} satellite peaks change, as we scan the phases, we can infer the relationship between the two free phases.  To have $P_{jk} = 0$ we need $\{\Phi_{jk}^R, \Phi_{jk}^L,\Phi_{jk}^R +\Phi_{jk}^L\} = \pm 2\pi /3$. If we parameterise this by just one of these phases, setting  $\Phi_{jk}^L =  n \Phi_{jk}^R$,  we can consider $\{\Phi_{jk}^R, n \Phi_{jk}^R, (n+1) \Phi_{jk}^R \} = \pm 2\pi /3$. As we see, there is a large family of possible solutions. To observe these effects we need to vary both of the phases at the same time and in a manner that satisfies these constraints.



\begin{figure}[!h]\label{fig:mphases}
\includegraphics[height=.22\textheight]{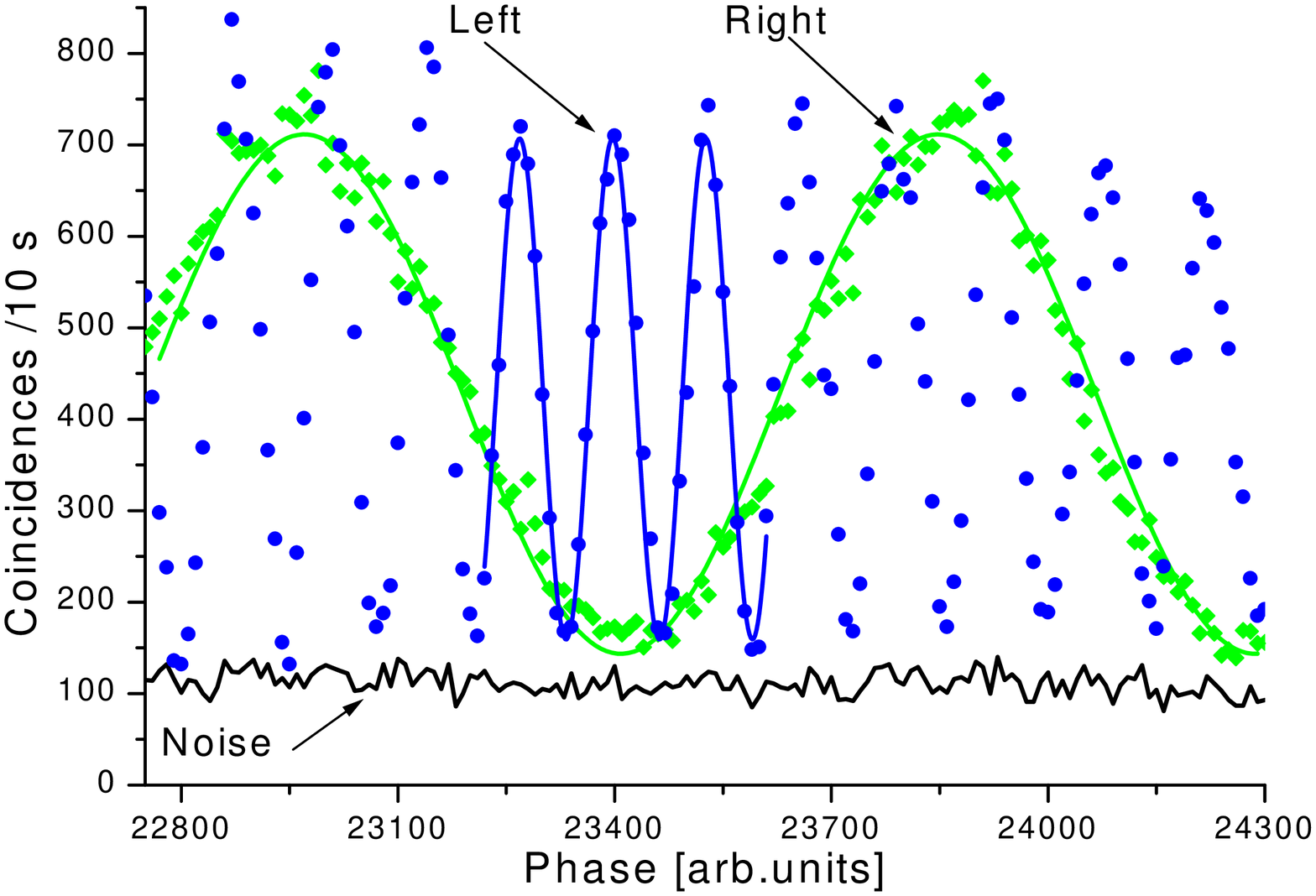}
\includegraphics[height=.22\textheight]{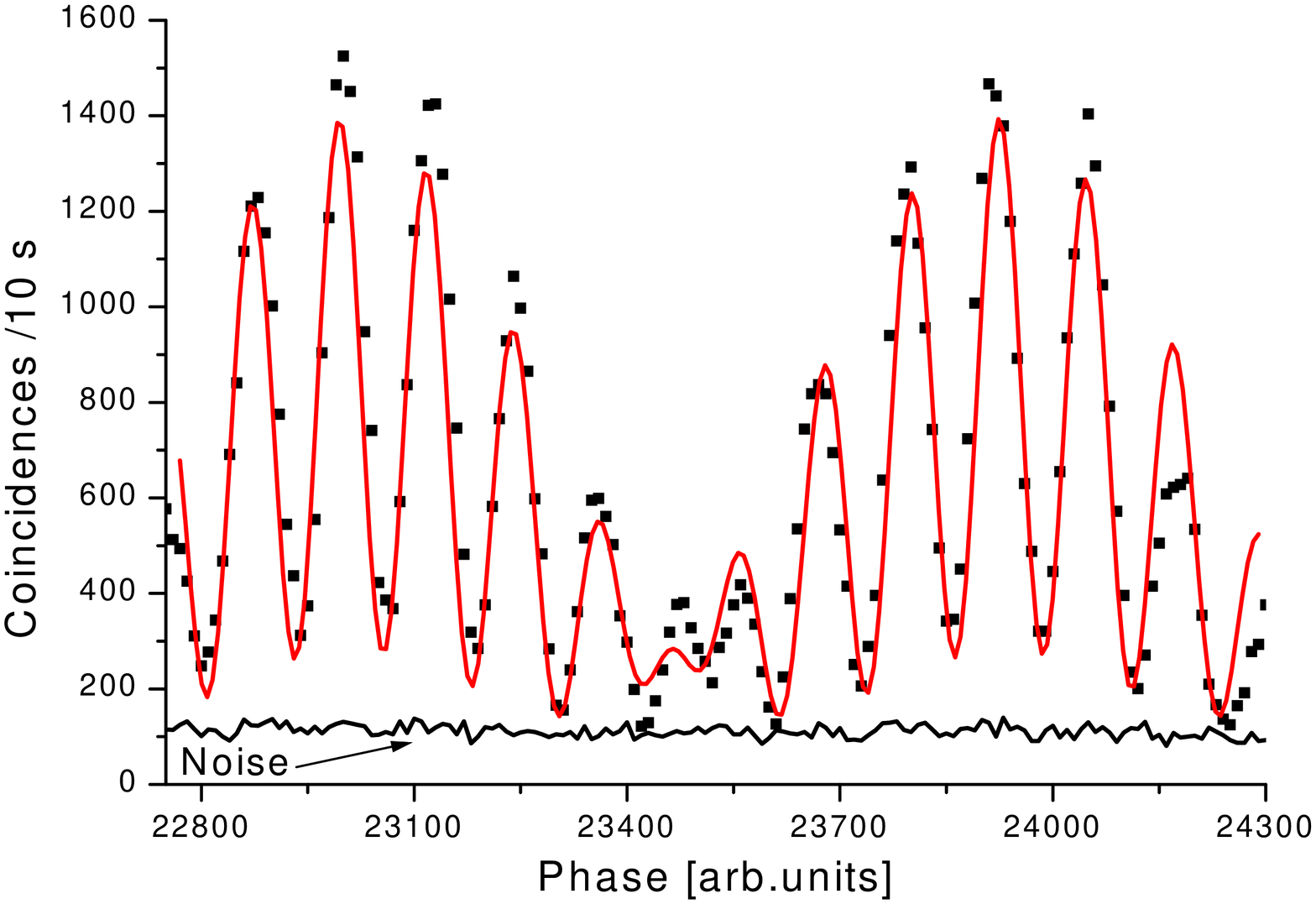}
\caption{(left) The coincidence counts for the {\it left} and {\it right} satellite peaks. The phase of the {\it left} peak is varying  $\approx 7$ times faster than phase of the {\it right}  peak.  (right) The coincidence counts, recorded simultaneously, for the central peak.}
\end{figure}

To illustrate let us consider a couple of examples. First, we have shown in Fig.\ref{fig:mphases} the case where the {\it left} phase is varying $n \approx 7$ times faster than the  {\it right} phase. One can easily verify that this satisfies the constraints we introduced for  $P_{jk} = 0$. 
The high visibility of the satellite peaks is reflected in that of the central peak where we see that the minimum  approaches the noise level. What we also see, on the right of Fig.\ref{fig:mphases} (central peak), is the beating type effect between the two phases. When the visibility is not high, in the central peak,  these satellite peaks allow us to determine whether this is due to the phase alignment or with standard misalignment and distinguishability problems.  From a practical perspective this significantly facilitates the alignment of the interferometers.

\begin{figure}[!h]\label{fig:2phases}
\includegraphics[height=.22\textheight]{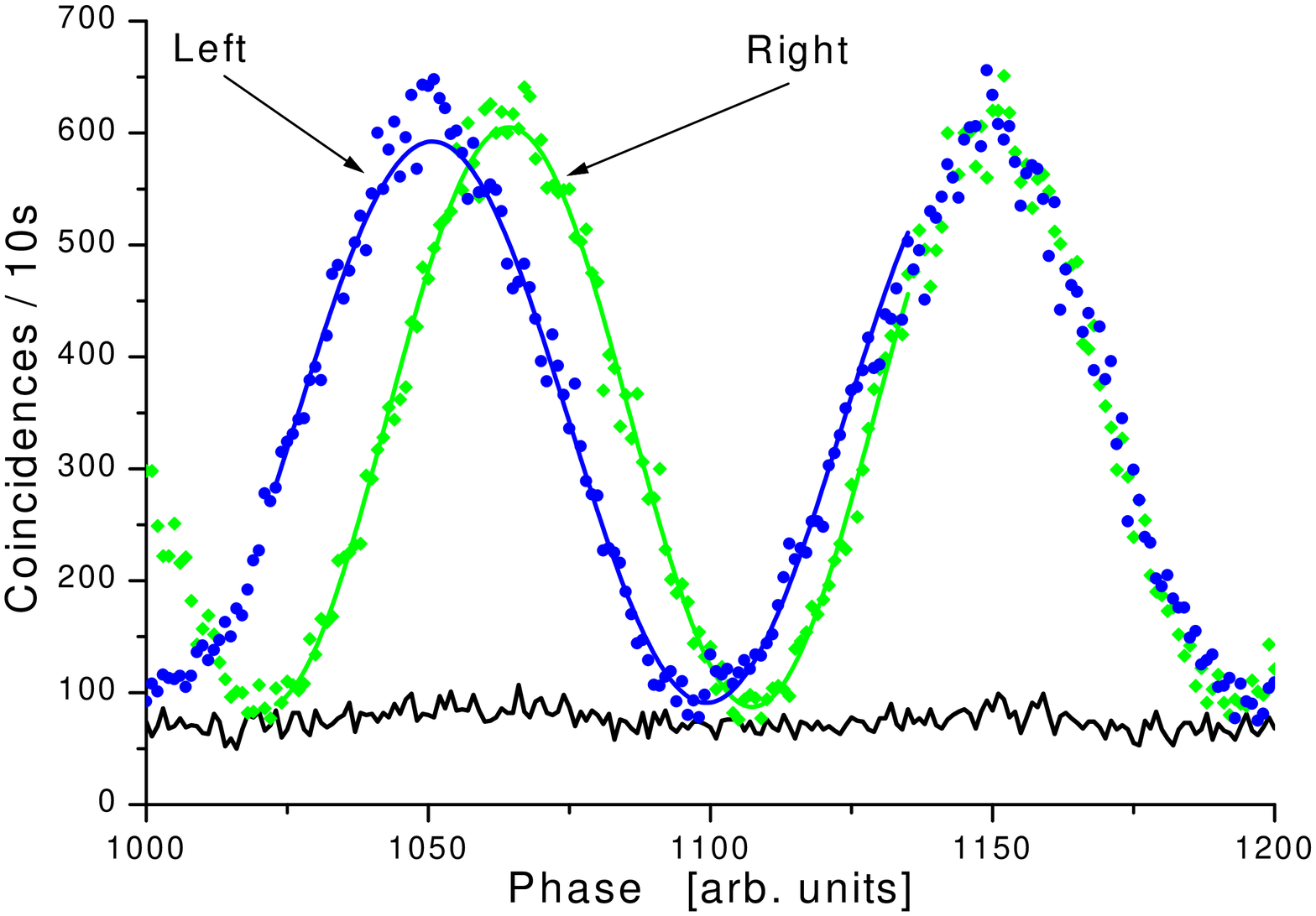}
\includegraphics[height=.22\textheight]{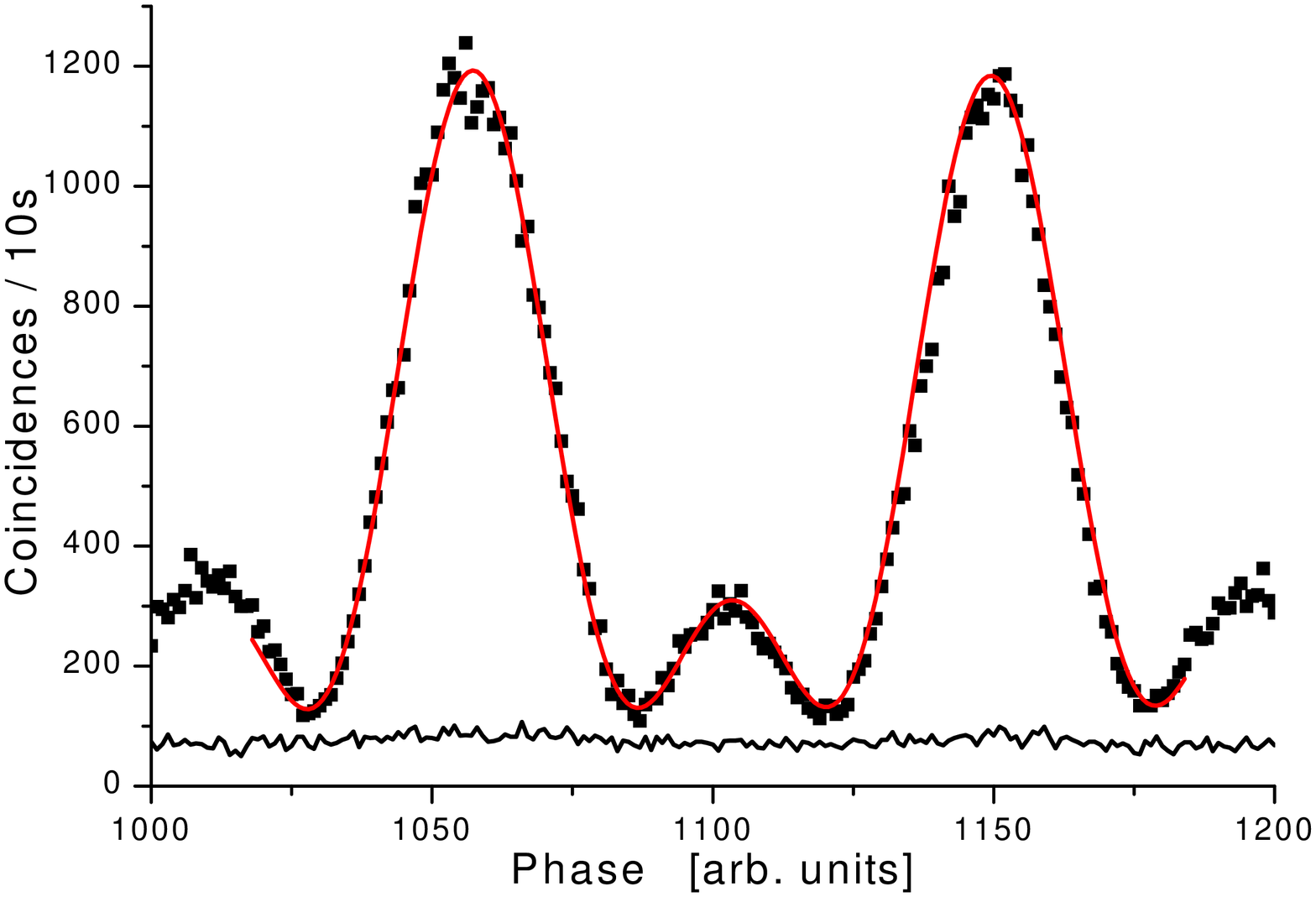}
\caption{Coincidence rates as in Fig.\ref{fig:mphases} with the {\it left} and {\it right}  phases  varying at almost the same rate, $n \approx 1$. }
\end{figure}

If we consider a second example we find that the simplest relationship is when both left and right vary at the same rate, $n \approx 1$, as depicted in Fig.\ref{fig:2phases}.
It turns out that this is exactly the same requirement as that needed to satisfy one of the symmetry constraints for the violation of the Bell inequality. The Bell test results and the relationship between the visibility and the CGLMP-Bell inequality \cite{Collins02a} are derived in \cite{Thew04b}.




Having a source of entangled qutrits that are capable of violating a Bell inequality we may ask what we can do with it. Proposals have been made for performing quantum key distribution (QKD) with high-dimensional systems \cite{Bechmann00a}. The setup illustrated in Fig.\ref{fig:expschematic} can be easily interpreted as a qutrit QKD set-up by associating the photon pair source with Alice, as denoted by the dashed square. We thus have a source that can be seen to transmit {\it heralded} qutrits from Alice to Bob in states determined by Alice's choice of phase setting \cite{Heralded}.  In  \cite{Bechmann00a} there are four bases introduced, the computational basis, $|0\rangle, |1\rangle, |2\rangle$, and three superposition bases. In this set-up we cannot use the computational basis without introducing switches at the input of the interferometers.  The simplest, and perhaps most naive, advantage one gains from using high-dimensional schemes for QKD is that one sends {\it dits} of information, which, from an information theoretic perspective, implies more information ($d$-dimensions $\rightarrow \log_2 d $ bits). The security bounds also show significant improvement with increased dimensioins. Cerf {\it et al} \cite{Cerf02a} studied both 2-basis and $d+1$ basis protocols, considering individual and coherent attacks, and found a significant security increase scaling with the dimensions. Specifically for qutrits, with individual attacks,  the 2-basis protocol is slightly lower, 21.13\%,  than the $d+1 (=4)$-basis, 22.67\%, whilst  for the coherent attacks the 11\% Shor-Preskill error bound for qubits goes up to almost 16\%. We see that there are definite advantages to be gained here, however, we are currently constrained by technical complications. 

There are at least two significant problems for high-dimensional schemes such as this. The first is that we have nine detectors that will all contribute to the QBER. Preliminary studies on the security and transmission implications due to detector noise have been made \cite{Bourennane01a} but this is difficult to generalise as the results are both system and protocol dependent. A more fundamental problem here is the way we count "successful events", i.e. the transmission of a qutrit. We only look, in coincidence, at the events where the photons took the same path in each interferometer. This means that whenever the photons take different paths to each other they are discarded. This amounts to keeping only a third of the possible signal, even before one gets to basis reconciliation. Currently, qubit cryptographic protocols are quite capable of operation below the 11\% error limit, thus, the main benefit is due to transferring dits and not bits.


Quantum coin tossing is another protocol where it has been found that qutrits are advantageous \cite{Ambainis02a,Molina-Terriza04a}.  If we consider the states associated with the satellite peaks and assign  $\{s, m, l\}_A$ with $\{0, 1, 2\}_A$ and $\{s, m, l\}_B$ with $\{1, 0, 2\}_B$, we can rewrite  Eq.\ref{eqLRphasestates}, with appropriate choices of phases, as 
\begin{eqnarray}
|\psi_{00}^L\rangle =  |11\rangle \pm |20\rangle_{AB}  \hspace{2cm}  |\psi_{00}^R\rangle = |00\rangle \pm |12\rangle_{AB}   . \label{eq:cointoss}
\end{eqnarray}
Thus if Alice has a detection in the left (right) peak, projecting onto $|1\rangle \pm |2\rangle $   ($|0\rangle \pm |1\rangle$), she prepares the state $|0\rangle \pm |1\rangle $   ($|0\rangle \pm |2\rangle$)  to send to Bob as the proposed protocol requires. Notice that the choice of the basis, {\it left} or {\it right},  is made randomly by the photon. Unfortunately, from a practical perspective again, the security to ensure that no one is cheating can be lost due to detector noise and efficiency.  Barrett {\it et al.} \cite{Barrett04a} have discussed the related problem of loss in coin tossing and bit commitment protocols  and shown that in the presence of noise a single coin-toss is not possible but one can generate secure bit-strings.  Cheating is possible as one of the parties can take advantage of the statistical possibility of there being a photon and not detecting it, or having no photon and accidental detection, etc. This discussion is independent of the dimension, however a more practical analysis by  Langford {\it et al.} \cite{Langford04a}   also showed that any security gains obtained by using qutrits over qubits was highly sensitive to the fidelity of the prepared states.

In conclusion, we have given the results for a Bell-test performed using the entangled qutrit source presented here and shown some of the control techniques that can be incorporated. We have also put some practical perspective on some of the often cited benefits of high-dimensional systems via two specific examples, quantum key distribution and coin tossing. One protocol that was of great experimental motivation, that we did not discuss here, is the Byzantine Agreement problem. This is classically impossible, however an elegant solution was found requiring three entangled qutrits  \cite{Fitzi01a}. Unfortunately, it was recently shown that it could in fact be achieved with a simple QKD arrangement \cite{Iblisdir04a}. What this leaves us with are some very interesting and useful high-dimensional entanglement sources that are providing an excellent  test-bed for fundamental ideas. From  a quantum information perspective however we are waiting for  ideas that really exploit the complexity that these systems have to offer.


 This project is financed by the Swiss NCCR "Quantum Photonics" and the EU IST-FET project RamboQ.






\begin{thebibliography}{00}
\bibitem{Thew04b} R.~T.~Thew, A. Ac\'{i}n, H. Zbinden, and N. Gisin, Phys. Rev. Lett. {\bf 93}, 010503 (2004).
\bibitem{Bechmann00a} H.~Bechmann-Pasquinucci and A.~Peres, Phys. Rev. Lett. {\bf 85}, 3313 (2000).
\bibitem{Cerf02a} N.~J.~Cerf {\it et al.}, Phys. Rev. Lett. {\bf 88}, 127902 (2002).
\bibitem{Ambainis02a} A.~Ambainis, Proc. STOC 01, 134 (2001).
\bibitem{Collins02a} D.~Collins {\it et al}., Phys. Rev. Lett. {\bf 88}, 040404 (2002).
\bibitem{Massar02a} S.~Massar, Phys. Rev. A {\bf 65}, 032121 (2002).

\bibitem{QutritsExp}J.~C.~Howell, A.~Lamas-Linares, and D.~Bouwmeester, Phys. Rev. Lett. {\bf 88}, 030401 (2002), A.~Vaziri, G.~Weihs and A.~Zeilinger, Phys. Rev. Lett. {\bf 89}, 240401 (2002), H.~de Riedmatten  {\it et al.}, Phys. Rev. A {\bf 69}, 050304 (2004).
\bibitem{Langford04a} N.~K.~Langford  {\it et al.}, Phys. Rev. Lett. {\bf 93}, 053601 (2004).



\bibitem{Thew04a}  R.Thew, A. Ac\'{i}n, H. Zbinden and N. Gisin,  Quant. Inf. Comp. {\bf 4}, 93 (2004). 
\bibitem{Franson89a}J.~D.~Franson, Phys. Rev. Lett. {\bf 62}, 2205 (1989).
\bibitem{Tittel99a} W.~Tittel, J.~Brendel, N.~Gisin, and H.~Zbinden, Phys. Rev. Lett. {\bf 81}, 3563 (1999).
\bibitem{Thew02a} R.~T.~Thew {\it et al}, Phys. Rev. A {\bf 66}, 062304 (2002).
\bibitem{Tanzilli01a} S.~Tanzilli {\it et al}, Elect. Lett. {\bf 37}, 28 (2001).
\bibitem{Zukowski97a} M.~Zukowski, A.~Zeilinger and M.~A.~Horne, Phys. Rev. A {\bf 55}, 2564 (1997).

\bibitem{Heralded} G.~Molina-Terriza,  {\it et al}, Phys. Rev. Lett. {\bf 92}, 167903 (2004), S.~Fasel  {\it et al}, quant-ph/0408136 (2004).


\bibitem{Bourennane01a} M.~Bourennane  {\it et al}, J. Phys. A {\bf 35}, 10065 (2002).
\bibitem{Molina-Terriza04a} G.~Molina-Terriza, A.~Vaziri, R.~Ursin, A. Zeilinger, quant-ph/0404027 (2004).
\bibitem{Barrett04a}  J.~Barrett and S.~Massar, Phys. Rev. A  {\bf 69}, 022322 (2004).
\bibitem{Fitzi01a} M.~Fitzi, N.~Gisin and U.~Maurer, Phys. Rev. Lett. {\bf 87}, 217901 (2001).
\bibitem{Iblisdir04a} S.~Iblisdir and N.~Gisin, To be published in Phys. Rev. A, quant-ph/0405167 (2004).
\end{thebibliography}

\IfFileExists{\jobname.bbl}{}
 {\typeout{}
  \typeout{******************************************}
  \typeout{** Please run "bibtex \jobname" to optain}
  \typeout{** the bibliography and then re-run LaTeX}
  \typeout{** twice to fix the references!}
  \typeout{******************************************}
  \typeout{}
 }

\end{document}